\documentstyle[emulateapj] {article}
\newcommand{\ts}{\thinspace}
 
\lefthead{EVANS ET AL.}
\righthead{MOLECULAR GAS IN 3C 293}

\begin{document}

\title{Molecular Gas in 3C 293: The First Detection of CO Emission 
and Absorption in an FR II Radio Galaxy}

\author{A. S. Evans}
\affil{Division of Physics, Math, \& Astronomy MS 105-24, California
Institute of Technology, Pasadena, CA 91125; ase@astro.caltech.edu}

\author{D. B. Sanders \& J. A. Surace\altaffilmark{1}}
\affil{Institute for Astronomy, 2680 Woodlawn Drive, Honolulu, HI 96822;
sanders@ifa.hawaii.edu; jason@ipac.caltech.edu}

\and 

\author{J. M. Mazzarella}
\affil{IPAC, MS 100-22, California Institute of Technology, Jet Propulsion
Laboratory, Pasadena, CA 91125; mazz@ipac.caltech.edu}

\altaffiltext{1}{Present Address:  IPAC, MS 100-22, California Institute
of Technology, Jet Propulsion Laboratory, Pasadena, CA 91125}

\begin{abstract}

The first detection of CO emission in a Fanaroff-Riley Class II (i.e.,
edge-brightened radio morphology) radio galaxy is presented.
Multiwavelength (0.36-2.17 $\mu$m) imaging of 3C 293 shows it to be a disk
galaxy with an optical jet or tidal tail extending towards what appears to
be a companion galaxy 28 kpc away via a low surface brightness envelope.
The molecular gas appears to be distributed in an asymmetric disk rotating
around an unresolved continuum source, which is presumably emission from
the AGN.  A narrow ($\Delta v_{\rm abs} \sim 60$ km s$^{-1}$) absorption
feature is also observed in the CO spectrum and is coincident with the
continuum source.  Using the standard CO conversion factor, the molecular
gas (H$_2$) mass is calculated to be 1.5$\times10^{10}$ M$_\odot$, several
times the molecular gas mass of the Milky Way.  The high concentration of
molecular gas within the central 3 kpc of 3C 293, combined with the
multiwavelength morphological peculiarities, support the idea that the
radio activity has been triggered by a gas-rich galaxy-galaxy interaction
or merger event.

\end{abstract}


\section{Introduction}

There exists substantial evidence that galaxy interactions or mergers are
the trigger for the nuclear activity in radio galaxies.  Optical imaging
surveys by Heckman et al. (1986) and Smith \& Heckman (1989 a,b) have
shown that a significant fraction of low-redshift powerful ($P_{\rm 408
MHz} \gtrsim 3\times10^{25}$ W Hz$^{-1}$) radio galaxies possess
morphological peculiarities (e.g., tails, fans, bridges, and dust lanes)
commonly associated with the collisions of galaxies.  A significant
fraction of low-redshift radio galaxies are also known to be luminous
far-infrared sources (Golombek, Miley, \& Neugebauer 1988). In many cases,
the shape of the spectral energy distributions (SEDs) at mid and
far-infrared wavelengths are consistent with thermal emission from dust
heated by young massive stars and/or the AGN.

Improvements in millimeter receiver technology and the availability of
moderate size (i.e., total collecting area greater than 500 square meters)
millimeter arrays has made it possible to detect and spatially map
star-forming molecular gas in radio galaxies.  Molecular gas, which forms
on the surface of dust grains, is of particular interest in active
galaxies because it is also a likely source of fuel for the central engine
during the initial phases of the merger.  To date, several low-redshift
radio galaxies have been unambiguously detected with single-dish
telescopes, revealing substantial quantities ($1\times10^9 -
5\times10^{10}$ M$_\odot$) of molecular gas (Phillips et al. 1987;
Mirabel, Sanders, \& Kaz\`{e}s 1989; Mazzarella et al. 1993; Evans 1998;
Evans et al.  1999). To date, however, detections of CO emission have been
limited to radio compact and Fanaroff-Riley I (edge-darkened radio
morphology:  Fanaroff \& Riley 1974) radio galaxies, the latter of which
tend to have relatively weak radio power.

In this paper, the first detection of CO emission and absorption in a
Fanaroff-Riley II (edge-brightened radio morphology) radio
galaxy, 3C 293 ($P_{\rm 408 MHz} \sim 4.5\times10^{25}$ W Hz$^{-1}$), is
presented. The galaxy possesses optical morphological peculiarities and
extended radio and CO emission, making it an ideal source for studying the
relationship between emission in radio galaxies at multiple wavelengths.

The paper is divided into five sections. Section 2 is a discussion of the
optical, near-infrared, and millimeter observations of 3C 293. The data
reduction methods and molecular gas mass calculations are summarized in \S
3. Section 4 contains an estimate of the dynamical mass of 3C 293 based on
the CO data, a discussion of the 2.7 mm continuum emission and CO
absorption, and concludes with a detailed comparison of the morphologies
of the galaxy at optical, near-infrared, millimeter, and radio
wavelengths.  Section 5 summarizes the results. 

Throughout this paper, we adopt H$_0 = 75$ km s$^{-1}$ Mpc$^{-1}$ and $q_0
= 0.0$.  Thus, for a source at a redshift of 0.045, 815 pc subtends
1$\arcsec$ in the sky plane.

\section{Observations}

The interpretation of the CO data presented in this paper has benefited
from observations of the galaxy at other wavelengths. Below, the
ground-based U$^{\prime}$, B, I, and K$^{\prime}$-band imaging
observations, as well as the millimeter observations are
summarized.\footnote{The UH U$^{\prime}$ and K$^{\prime}$-band filters
have central wavelengths of 3410\AA$~$ and 21700\AA, respectively.}
Additional radio data and HST data have also been obtained courtesy of J.
P. Leahy and the Hubble Space Telescope (HST) Archive.

\subsection{Ground-based Imaging}

Ground-based imaging observations of 3C 293 were made at the UH 2.2m
Telescope.  The U$^{\prime}$, B, and I-band images were obtained on 1998
March 25 using the Orbit Semiconductor 2048$\times$2048 CCD camera. The
original scale is $0.09\arcsec$/pixels (Wainscoat 1996), but the CCD was
read out with $2\times2$ pixel binning. Four dithered exposures were
taken, with integration times of 480, 360, and 360 seconds each for
U$^{\prime}$, B, and I, respectively.  Near-infrared, K$^{\prime}$
observations of 3C 293 were also obtained on 1996 April 24 using the UH
QUick Infrared Camera (QUIRC:  Hodapp et al. 1996), which consists of a
1024$\times$1024 pixel HgCdTe Astronomical Wide Area Infrared Imaging
(HAWAII) array. The near-infrared observations were done at f/10,
providing a field of view of $3\arcmin\times3\arcmin$. Five dithered
exposures were taken, each with an integration time of 180 seconds.

\subsection{CO Spectroscopy}

The initial millimeter observations of  3C 293 were made with the
NRAO\footnote{The NRAO is a facility of the National Science Foundation
operated under cooperative agreement by Associated Universities, Inc.} 12m
Telescope on 1996 January 24.  The telescope was configured with two
$256\times2$ MHz channel filterbanks and dual polarization SIS
spectral-line receivers tuned to the frequency 110.35 GHz, corresponding
to a redshift of 0.045 (Sandage 1966; Burbidge 1967) for the CO($1\to0$)
emission line.  Observations were obtained using a nutating subreflector
with a chop rate of $\sim 1.25$ Hz. Six minute scans were taken, and a
calibration was done every other scan. Pointing was done on 3C 273 prior
to the observations, and the pointing was checked at the end of the
observations using Mars.  The total duration of the observations was 8.8
hours.

Follow-up aperture synthesis maps of CO($1\to0$) and 2.7 mm continuum
emission in 3C 293 were made with the Owens Valley Radio Observatory
(OVRO) Millimeter Array during three observing periods from 1997 September
to November.  The array consists of six 10.4m telescopes, and the longest
observed baseline was 242m. Each telescope was configured with
120$\times$4 MHz digital correlators. Observations done in the
low-resolution configuration (1997 September and October) provided a
$\sim4.0\arcsec$ (FWHM) synthesized beam with natural weighting, and
observations in the high-resolution configuration (1997 November) provided
a beam of $\sim2.5\arcsec$ (FWHM).  During the observations, the nearby
quasar HB89 1308+326 (1.27 Jy at 110 GHz; B1950.0 coordinates
$13^h08^m07.56^s +32^{\rm o}36^{\prime}40.23^{\prime\prime}$) was observed
every 25 minutes to monitor phase and gain variations, and 3C 273 was
observed to determine the passband structure.  Finally, observations of
Uranus were made for absolute flux calibration.

\section{Data Reduction and Results}

\subsection{Imaging Data}

The U$^{\prime}$, B, and I-band data reduction was performed using IRAF.
The data reduction consisted of flatfielding individual images, scaling
each image to its median value to correct for offsets in individual
images, then shifting and median combining the images.  The final images
were then boxcar smoothed by $4\times4$ pixels.

Figure 1 shows two three-color images of 3C 293. In Figure 1a, 3C 293 is
presented in a linear stretch to show the highest surface brightness
features. Figure 1b shows the galaxy with a logarithm stretch, revealing
additional, extended low surface brightness structure.

The K$^{\prime}$-band data reduction was done in a similar manner to the
U$^{\prime}$, B, and I-band data reduction, except that after the median
level of each image was subtracted, the individual frames were averaged
together without spatial shifting using a min/max averaging routine. This
procedure removes the contribution of astronomical sources in the
individual frames to produce an object-free ``sky'' image.  This sky image
was then subtracted from the individual frames before they were shifted
and averaged.  The resultant wide-field K$^{\prime}$-band image is
discussed in \S 4.5.

\subsection{NRAO 12m Telescope Data}

The data reduction for the NRAO 12m telescope data were reduced using the
IRAM data reduction package CLASS.  The individual scans were checked for
baseline instabilities, then averaged together. The emission line was
observed to span the velocity range -500 to 350 km s$^{-1}$ (where 0 km
s$^{-1}$ corresponds to the systemic velocity of 3C 293), thus a linear
baseline was subtracted excluding this velocity range. The spectrum was
then smoothed to 43 km s$^{-1}$.

Figure 2 shows the CO(1$\to$0) spectrum. The emission line is moderately
broad -- the velocity width at half the maximum intensity, $\Delta v_{\rm
FWHM}$, of $\sim 400$ km s$^{-1}$. For comparison, the mean value for
infrared luminous galaxies is 250 km s$^{-1}$ (Sanders, Scoville, \&
Soifer 1991).  An absorption feature is also observed at the systemic
velocity of 3C 293 ($\sim 13500$ km s$^{-1}$), corresponding to the
velocity of the HI absorption feature associated with the galaxy (Baan \&
Haschick 1981).

The CO emission line has a intensity, $I_{\rm CO} = T_{\rm mb} \Delta v$,
of $2.0\pm 0.4$ K km s$^{-1}$, where $T_{\rm mb}$ is the main-beam
brightness temperature. Using a Kelvin-to-Jansky conversion factor of 25.2
Jy K$^{-1}$ (P. Jewell 1995, private communication), the emission-line
flux, $S_{\rm CO} \Delta v$, is 51$\pm 11$ Jy km s$^{-1}$.

\subsection{OVRO Data}

The OVRO data were reduced and calibrated using the standard Owens Valley
array program MMA (Scoville et al. 1992). The data were then exported to
the mapping program DIFMAP (Shepherd, Pearson, \& Taylor 1995). 

The resultant continuum and integrated intensity maps are shown in Figure
3.  The CO emission in 3C 293 is extended over a region 7$\arcsec$ (5.7
kpc) in diameter.  Four spectra have also been extracted, showing clearly
the narrow absorption feature ($\Delta v_{\rm abs} \sim 60$ km s$^{-1}$)
seen in the single-dish spectrum. In contrast to the CO emission, the
continuum emission is unresolved, and appears to be located at the center
of the molecular gas morphology. Both the synthesized maps of the CO
absorption (not shown) and the 2.7 mm continuum are spatially coincident,
indicating the presence of CO emitting gas along the line of sight to the
continuum source.

The total flux measured within a 7$\arcsec$ diameter aperture is 53$\pm
10$ Jy km s$^{-1}$, consistent with the flux obtained from the 12m in a
74$\arcsec$ beam.

\subsection{Molecular Gas Mass}

For a $q_0 = 0.0$ universe, the luminosity distance for a
source at a given redshift, $z$, is,
$$D_{\rm L} = cH^{-1}_0 z \left(1 + 0.5z \right)
~~[{\rm Mpc}].
\eqno(1)$$
Given the measured flux, 
$S_{\rm CO} \Delta v$,
the CO luminosity of a source at
redshift $z$ is,
$$L'_{\rm CO} = \left( {c^2 \over {2 k \nu^2_{\rm obs}}} \right)
S_{\rm CO} \Delta v D^2_{\rm L} (1 + z)^{-3}, \eqno(2)$$
where $c$ is the speed of light, $k$ is the Boltzmann constant,
and $\nu_{\rm obs}$ is the observed frequency.
For a luminosity distance expressed in units of Mpc, $L'_{\rm CO}$
can be written as,
$$L'_{\rm CO} = 2.4\times10^3
\left( S_{\rm CO} \Delta v \over {\rm Jy~km~s}^{-1} \right)
\left( D^2_{\rm L} \over {\rm Mpc^2} \right) (1 + z)^{-1}$$
$$~~~~~~~~~~~~~~~~~~~~~~~~
~~~~~~~~~~~~~~~[{\rm K~km~s}^{-1} {\rm~pc}^2]. \eqno(3)$$

For 3C{\ts}293, $S_{\rm CO} \Delta v = 51$ Jy km s$^{-1}$ and $D_{\rm L} =
18 0$ Mpc, thus $L'_{\rm CO} = 3.8\times10^9\ {\rm K~km~s}^{-1}
{\rm~pc}^2$.  To calculate the mass of molecular gas in 3C{\ts}293, a
reasonable assumption is made that the CO emission is optically thick and
thermalized, and that it originates in gravitationally bound molecular
clouds. Thus, the ratio of the H$_2$ mass and the CO luminosity is given
by, $\alpha = M($H$_2) / L^\prime_{\rm CO} \propto \sqrt {n({\rm H}_2)} /
T_{\rm b}$ M$_\odot$ (K km s$^{-1}$ pc$^2)^{-1}$, where $n($H$_2)$ and
$T_{\rm b}$ are the density of H$_2$ and brightness temperature for the
CO(1$\to$0) transition (Scoville \& Sanders 1987; Solomon, Downes, \&
Radford 1992).  Multitransition CO surveys of molecular clouds in the
Milky Way (e.g. Sanders et al. 1993), and in nearly starburst galaxies
(e.g. G\"{u}sten et al. 1993) have shown that hotter clouds tend of be
denser such that the density and temperature dependencies tend to cancel
each other. The variation in the value of $\alpha$ is approximately a
factor of 2 for a wide range of kinetic temperatures, gas densities, and
CO abundance (e.g. $\alpha = 2-5 M_{\odot}$ [K km s$^{-1}$ pc$^2]^{-1}$:
Radford, Solomon, \& Downes 1991) We adopt a value of 4 $M_{\odot}$ (K km
s$^{-1}$ pc$^2)^{-1}$ for $\alpha$, which is similar to the value
determined for the bulk of the molecular gas in the disk of the Milky Way
(Strong et al. 1988; Scoville \& Sanders 1987).\footnote{The equivalent
$X_{\rm CO} = n({\rm H}_2)/I_{\rm CO} = 2.5\times10^{20}$ cm$^{-2}$ (K km
s$^{-1}$)$^{-1}$.}  Thus, the molecular gas mass of 3C{\ts}293 is
$1.5\times10^{10}$ M$_{\odot}$, ($\sim${\ts}5 times the molecular gas mass
of the Milky Way), however, note that $M({\rm H}_2)$ could be as low as
$7.3\times10^{9}$ M$_{\odot}$ ($\sim 2.5$ times the Milky Way molecular
gas mass).  In addition, the concentration of molecular gas within the
inner 3 kpc of 3C 293 is $\sim$ 530 M$_\odot$ pc$^{-2}$. Such a molecular
gas concentration is 4-400 times greater than that of nearby early-type
spiral galaxies (e.g., Young \& Scoville 1991), but on the low end of
H$_2$ concentrations determined for a sample of $L_{\rm ir} > 10^{11}$
L$_\odot$ merging galaxies studied by Scoville et al. (1991) and Bryant
(1996).

\section{Discussion}

As mentioned in \S 1, several other radio galaxies have been detected in
CO by single-dish telescopes. Millimeter interferometry of these radio
galaxies is currently underway; for the present moment, the global
properties of 3C 293 will be briefly compared with these radio galaxies
before discussing 3C 293 in detail.

Table 1 summarizes the infrared and CO properties of 3C 293 relative to
other CO-luminous radio galaxies.  The galaxy 3C 293 is one of the most
molecular gas rich of the radio galaxies detected to date, but has a
relatively low infrared luminosity and low $L_{\rm ir}/L^{\prime}_{\rm
CO}$ ratio.  The ratio $L_{\rm ir}/L^{\prime}_{\rm CO}$ is commonly
referred to as the star formation ``efficiency''; if the infrared
luminosity is dominated by reprocessed light from young, massive stars,
then the ratio indicates how efficiently molecular gas is converted into
stars. In the case of 3C 293, a ratio of 8.9 indicates that the galaxy is
producing stars at the rate of quiescent spiral galaxies (e.g., Sanders \&
Mirabel 1996). All of the more infrared luminous radio galaxies have very
high $L_{\rm ir}/L^{\prime}_{\rm CO}$ ratios, but caution should be taken
when interpreting their infrared luminosities; in the most luminous
infrared galaxies, there is strong evidence that the infrared luminosity
is heavily contaminated by reprocessed AGN light (e.g., Evans et al.
1998a). We will briefly return to this discussion in \S 4.6.

\subsection{Optical Morphology}

Heckman et al. (1986) described 3C 293 as a primary galaxy connected to a
southwestern companion galaxy by a bridge of emission which fans into a
westward extending tidal tail beyond the companion.  While the images in
Figure 1 do not have a wide enough field of view to show the tidal tail
associated with the companion galaxy, Figure 1b clearly shows the low
surface brightness emission that appears to be connecting the two
galaxies. Indeed, the halo appears to completely envelope the primary
galaxy.

The high surface brightness morphology of the primary galaxy is clearly
that of a disk galaxy (Figure 1a). The bulge component has redder colors
than the disk component, and contains a prominent, warped dust lane (see
also van Breugel et al. 1984). The disk component also appears to be
warped, but its apparent morphology may simply be an artifact of dust
obscuration. It is not well understood what causes galaxy warps, but in
the case of 3C 293, the warp may result from gravitational interactions
with the companion galaxy, gas infall, and/or a misalignment between the
disk and halo (e.g. Binney 1992).

\subsection{CO Emission}

The CO distribution of 3C 293 appears to be asymmetric (Figure 3), and may
be subject to the same warping observed in the dust lanes and possibly the
large-scale disk.  As is clear from the four extracted spectra and CO
morphology, the molecular gas is distributed in a disk rotating around the
unresolved continuum source.  Taking a disk radius of 3.5$\arcsec$ (2.8
kpc) and a velocity width at half the maximum intensity of $\Delta v_{\rm
FWHM} \sim 400$ km s$^{-1}$, the dynamical mass is

$$M_{\rm dyn} \approx {r{\ts}\Delta v^2_{\rm FWHM} 
\over \beta G{\ts}{\rm sin}^2i} =
1.0\times10^{11} ~~~[({\rm sin}^2 i)^{-1} {\rm M}_{\odot}],
\eqno(4)$$
where $G$ is the gravitational constant, $i$ is the disk inclination, and
$\beta$ is approximately unity (e.g. Bryant \& Scoville 1996).
Thus the molecular gas mass is $\sim$ 10\% or less of the estimated
dynamical mass. The remainder of the mass within this radius consists of
stars and the central engine.

A molecular gas warp has been inferred in the radio galaxy Centaurus A by
Quillen et al. (1992). From models of their single-dish telescope maps,
Quillen et al. have speculated that the gas is in a triaxial potential. If
the molecular gas in 3C 293 is also in such a potential, equation 4 may
overestimate the true dynamical mass (i.e., if the gas is in elliptical
orbits instead of circular). Thus the molecular gas in the inner 2.8 kpc
may constitute more than 10\% of the total estimated gas mass.

\subsection{The 2.7 mm Continuum Emission}

What is the source of the 2.7 mm continuum emission? The continuum
emission, which has a flux density of 0.19 Jy, fits the extrapolated
$f_{\nu} \propto \nu^{-\alpha}$ power-law relation of the higher frequency
radio flux densities\footnote{The radio densities were obtain from data
compiled by the NASA Extragalactic Database (NED).}, indicating that the
2.7 mm continuum emission is nonthermal.

As a doublecheck, assume that the 2.7 mm emission is due to thermal
emission from dust, such as is the case for Arp 220. Arp 220, which is at
a distance of 77 Mpc, has a 2.7 mm flux density of 0.035 Jy (Scoville,
Yun, \& Bryant 1997).  Thus, the 2.7 mm luminosity of 3C 293 is 30 times
higher than that of Arp 220. If the assumption is made, as for Arp 220,
that the emission at $\lambda_0 > 200\mu$m is emanating from optically
thin ($\tau \sim 1$) dust radiating at a temperature of $\sim 40$K, the
dust mass is given by, $$M_{\rm dust} \approx {{S_{\rm obs}
D^2_L}\over{(1+z) \kappa_0 B(\nu_0,T)}},\eqno(5)$$ where $S_{\rm obs}$ is
the observed flux density, $\kappa_0$ is the rest-frequency mass
absorption coefficient with a value of 0.085 g$^{-1}$ cm$^2$ (i.e., 10
g$^{-1}$ cm$^2$ at 250 $\mu$m scaled to 2.7 mm:  Hildebrand 1983),
$B(\nu_0,T)$ is the rest-frequency value of the Planck function.  The
derived dust mass of 3C 293 is thus $2.0\times10^{10}$ M$_{\odot}$.
Further, if the dust in 3C 293 is radiating at a temperature of 15K, the
derived dust mass is a factor of 3 higher.  Given that the standard
gas-to-dust ratio is 100--200, and that the 2.7 mm continuum emission is
not spatially extended like the CO emission, the 2.7 mm continuum emission
is unlikely to be thermal emission emanating from dust associated with the
molecular gas.  Thus, the continuum emission most likely emanates from
processes directly associated with a supermassive nuclear black hole or
circumnuclear accretion disk.

\subsection{The Nature of the CO Absorption}

HI absorption features in several low-redshift radio galaxies have been
extensively studied (e.g. Baan \& Haschick 1981; Mirabel 1989; Conway \&
Blanco 1995).  Such studies have attempted to determine the location and
mass of HI within the host galaxies.  In 3C 293, the depth of the CO
absorption is shallower in the single-dish spectrum than in the OVRO
spectrum, indicating that beam dilution is masking the true absorption
depth. Given this, the optical depth, $\tau_{\rm CO}$, of 3C 293 is, $\tau
\gtrsim$ ln$(I_{\rm cont}/(I_{\rm cont} - \Delta I_{\rm abs})) \gtrsim
0.69$, where $I_{\rm cont}$ is the continuum emission flux density and
$\Delta I_{\rm abs}$ is the CO absorption depth.

While the velocity dispersions of individual absorbing clouds are probably
typical of giant molecular clouds (i.e., $\lesssim 10$ km s$^{-1}$), the
dispersion of the absorption feature ($\Delta v_{\rm abs} \sim$ 60 km
s$^{-1}$) may be due to the motion of the clouds around the nucleus of the
galaxy. Assuming this is the case, 
the absorption occurs in clouds at a distance of
$r_{\rm cloud} = M_{\rm BH} G / \Delta v_{\rm abs}^2 \sim 120
(M_{(<r)}/10^8 {\rm M}_\odot)$ pc from 
the continuum source, where $M_{(<r)}$ is the
stellar/gaseous/black hole mass interior to the absorbing clouds and G is
the gravitational constant.  This is comparable to the metric size of the
dust torus surrounding the nucleus of the nearby FR{\ts}II radio galaxy
3C{\ts}270 (NGC 4261), detected using the HST (Jaffe et al. 1996).

Is the velocity dispersion of the absorbing clouds really due to
circumnuclear clouds in circular orbits?  Similar stellar dispersion
velocities have been observed in several galaxies believed to contain
quiescent black holes (i.e., galaxies that may have once been radio
galaxies or quasars):  in the dwarf spheroidal M32, which is believed to
possess a $2\times10^6$ M$_{\odot}$ black hole, the velocity dispersion is
60 km s$^{-1}$ and increases to $\sim 90$ km s$^{-1}$ within the central
1$\arcsec$ (4 pc: e.g., Kormendy \& Richstone 1995). The elliptical galaxy
NGC 3377, believed to harbor a $2\times10^8$ M$_{\odot}$ black hole, has a
velocity dispersion of 90 km s$^{-1}$ which increases to $\sim 158$ km
s$^{-1}$ within the central 1$\arcsec$ (48 pc: Kormendy et al. 1998).
However, if the dynamical mass within the inner 2.8 kpc of 3C 293 is on
the order of $1\times10^{11}$ M$_{\odot}$, 3C 293 as a whole may have a
mass comparable to a massive elliptical galaxy ($\gtrsim 3\times10^{11}$
M$_{\odot}$). Within 100 pc of the nucleus, the dispersion may actually be
$\sim 150-250$ km s$^{-1}$, similar to those observed for NGC 3115 and NGC
4594 (The Sombrero Galaxy: e.g. Kormendy \& Richstone 1995).  The CO
absorption may then be occuring in clouds in noncircular motion well
outside of the circumnuclear region, such as is believed to be the case
for some of the clouds responsible for the molecular absorption observed
in Centaurus A (Wiklind \& Combes 1997).

\subsection{Comparison Between the CO Distribution of Radio Morphology}

In Figure 4, the large-scale radio emission of 3C 293 (Leahy, Pooley, \&
Riley 1986) is superimposed on the wide-field K$^{\prime}$-band image.
The large jets appear to be oriented nearly perpendicular to the major
axis of the primary galaxy. Based on the major axis of the CO distribution
(Figure 3) and the large scale structure, one might naively assume that
the radio jets have simply escaped along the path of least resistance, the
path perpendicular to the molecular gas and stellar distribution. Such a
scenario would be consistent with the idea that the central engine is fed
by a molecular torus/accretion disk and that the radio jets escape along
the axis perpendicular to the accretion disk (e.g. Blandford 1984).

The structure in the inner few arcseconds, however, complicates matters:
Figure 4a shows a high-resolution 5 GHz MERLIN map (Akujor et al.  1996)
superimposed on both the high-resolution CO map and an archival HST 7000
\AA$~$ of 3C 293. The radio map has been registered with the CO map by
assuming that the core of the 5 GHz emission is coincident with the 2.7 mm
continuum emission (the coordinates of the 5 GHz core and the 2.7 mm
continuum emission differ by $<0.3\arcsec$, which is much less than the
resolution of the OVRO data).  The registration of the MERLIN map with the
HST image was done by first aligning the radio knots with features
observed in a recently obtained near-infrared image of 3C 293 taken with
the Near-Infrared Camera and MultiObject Spectrometer (NICMOS) aboard HST
(Evans et al. 1998b).  Such an alignment places the core of the 5 GHz
emission on what appears to be the nucleus of the galaxy. Note that the
nuclear dust lane observed in Figure 1 breaks up into multiple dust lanes
in the HST image, making the nuclear region of 3C 293 resemble that of
Centaurus A, which is dissected by a prominent warped dust lane (Sandage
1961).

The axis of the inner radio emission that extends 2.5$\arcsec$ (2.0 kpc)
from the nucleus is rotated $\sim 30^{\rm o}$ relative to the outer 1.5
GHz radio emission, which itself spans 200$\arcsec$ (160 kpc) from the tip
of the bright NW lobe to the tip of the fainter SE lobe. Indeed, the twist
of the eastern jet is apparent in the large-scale emission, which curves
from the core towards the southern lobe.  Such a jet morphology can be
caused by {\it i)} the redirection of the jets from a glancing impact with
a molecular cloud or from a sudden change in the density gradient
(e.g. van Breugel et al. 1984), {\it
ii)} by a merger event or interaction, causing a warp in the galaxy and
thus shifting the radio axis (as in IRAS P09104+4109: Hines et al. 1998), 
or {\it iii)} by a precessing radio jet.
The first scenario is supported by the fact that the eastern jet appears
to twist northward after passing the 7000\AA$~$ off-nuclear knot, then
southward after passing the centroid of the western CO component.  At
larger radii, the galactic density gradient would drive the large-scale
radio emission to be perpendicular to the molecular and stellar
distribution of the galaxy.  However, it is difficult to reconcile this
scenario with the relativistic velocity of the large-scale radio jets, as
implied by the apparent Doppler boosting (dimming) of the
approaching NW (receding SE) jets; an impact with molecular clouds would
undoubtedly cause the inner jet to lose substantial amounts of kinetic
energy.  The second scenario is supported by the morphologies of the
primary and companion galaxies (Figure 1; Heckman et al. 1986).  Given
this, the inner radio jets may represent a more recent outburst triggered
by a galaxy-galaxy interaction.  Following Akujor et al. (1996), if it is
assumed that the jet propagation speed is $0.1c$, the radio outburst
resulting in the large scale structure occured $5\times10^6$ years ago,
and the companion outburst was produced only $5\times10^4$ years ago.  The
third scenario is related somewhat to the second; the creation of a radio
jet in a dynamically unrelaxed environment could result in the observed
precession, and thus the high-resolution MERLIN map reflects the present
pole axis of the supermassive black hole.

\vskip 0.3in
\subsection{The Interaction/Merger Status of 3C 293}

Much of the interpretation of the dynamically state of 3C 293 is dependent
on whether or not the companion galaxy is the trigger for the strong
interaction/merger.  There is indeed strong morphological evidence that
the companion is associated with 3C 293 (\S 4.1), but to our knowledge,
there exists no published redshift of the companion.  However, the
companion appears to be a galaxy and not a foreground star; the measured
FWHM of the companion in the 2.17 $\micron$ image is 2.1$\arcsec$, as
compared with 1.2$\arcsec$ for stars in the field.  

{}From the morphologies of 3C 293 and the companion, it might at first
appear that the companion penetrated through the nuclear region of 3C 293
approximately (28 kpc / 250 km s$^{-1}) \sim$ $1\times10^8$ years ago,
leaving a trail of optical/near-infrared debris (e.g., Figures 1 and 4).
Alternatively, the bridge of optical/near-infrared emission between 3C 293
and the companion may be a nonaxisymmetric tidal feature viewed nearly
edge-on, created as the orbit of the companion decays via dynamical
friction (see simulations by Hernquist \& Mihos 1995).  However, a
determination of the relative masses of the two galaxies suggests that the
companion is not massive enough to have caused the morphological
disturbances observed in 3C 293 (Figure 4a). The relative masses of the
two galaxies can be estimated from the fact that the spectral energy
distribution of supergiant stars peaks at $\sim$1.6 $\mu$m, and thus the
observed 2.17 $\mu$m luminosity is correlated with the mass of the
galaxy.  The ratio of companion mass to the mass of 3C 293 is therefore
estimated to be $\sim0.03$.  Note that this ratio does not include light
from the nucleus of 3C 293, which may emanate from the active galactic
nucleus.

Given that interactions with the companion are an unlikely trigger for the
present state of 3C 293, the linear feature extending from the host galaxy
is most likely a tidal remnant from a merger event occuring more than
10$^9$ years ago (see simulations by Barnes \& Hernquist 1992, 1996).
Similar one-sided tidal features are observed in advanced mergers such as
Mrk 273 (e.g. Mazzarella \& Boroson 1993). A major difference between 3C
293 and advanced mergers such as Mrk 273, Arp 220, and the radio galaxy 3C
120 is that 3C 293 has a relatively low $L_{\rm ir} / L^{\prime}_{\rm CO}$
ratio.  This, however, may simply indicate that the star formation and/or
AGN activity in 3C 293 that is typically responsible for boosting the
infrared luminosity is waning.

The instabilities resulting from the merger have undoubtedly induced a
loss of angular momentum in the molecular disk, causing a fraction of the
gas to fall inward, ultimately serving as fuel for the nuclear activity.
Eventually, the stellar distribution will dynamically relax into that of
an elliptical or an early-type spiral galaxy, possibly making 3C 293
resemble more classical FR II radio galaxies such as Cygnus A.

\section{Summary}

In this paper, the first detection of CO emission in an FR II Class radio
galaxy, the galaxy 3C 293, has been presented. The analysis of the CO data
has benefited from data taken at a wide range of wavelengths. The
following conclusions are drawn:

1. The optical morphology of 3C 293 shows a primary and what appears to be
a companion galaxy enveloped by a low surface brightness halo, consistent
with previous optical images of the galaxy. In addition, the primary
galaxy has a warped dust lane and a possible warped disk, most likely
resulting from an interaction/merger, gas infall, and/or misalignment
between the disk and halo of the primary galaxy.

2. The CO emission line is relatively broad ($\Delta v_{\rm FWHM} \sim
400$ km s$^{-1}$), and contains a $\Delta v_{\rm abs} \sim$ 60 km s$^{-1}$
absorption feature at the same velocity as the HI absorption feature
previously observed.  Using the standard CO-to-H$_2$ mass conversion
factor, the inferred molecular gas mass of 3C 293 is $1.5\times10^{10}$
M$_{\odot}$.

3. The CO emission in 3C 293 is observed to be extended over a 7$\arcsec$
(5.7 kpc) diameter region, and the H$_2$ concentration within the inner 3
kpc is $\sim$ 530 M$_{\odot}$ pc$^{-2}$.  An unresolved, 2.7 mm continuum
source is also detected, which is spatially coincident with the systemic
absorption feature.  The morphology and kinematics of the CO emission are
consistent with a disk of molecular gas rotating around the continuum
source.

4. The flux density of the 2.7 mm continuum source (0.19 Jy) is consistent
with a power-law extrapolation of the radiowave flux densities.  This, and
the fact that the continuum emission is not coincident with the CO
emission, indicates that the 2.7 mm continuum emission is nonthermal in
nature, and that it most likely emanates from a circumnuclear accretion
disk around a supermassive black hole, and not from dust associated with
the molecular gas.

5. The radio jets of 3C 293 appear to twist by about 30${\rm ^o}$ from the
inner 10$\arcsec$ to the radio lobes. The cause of the twist is either due
to the effects of a varying interstellar density gradient, or precession,
both of which are connected with a strong interaction or merger event in
the recent history of 3C 293.

Based on the above conclusions, the observed properties of 3C 293 result
from a strong interaction or merger event.  The resultant instabilities
have most likely induced gas in the inner molecular disk to lose angular
momentum and fall into the center of the gravitational potential well.
Such a scenario provides a plausible explanation of how the central
engines of radio galaxies are fueled.

\acknowledgements

A.S.E. thanks A. Readhead, M. Shepherd, N. Scoville, and T. Pearson for
many useful discussions and J. P. Leahy for providing the 1.5 and 5 GHz
radio maps. We thank the staffs of the NRAO 12m telescope, OVRO, and the
UH 2.2m telescope for their assistance. We also thank the referee for many
useful suggestions.  A.S.E. and D.B.S. are supported in part by NASA
grants NAG5-3042 and NAG5-3370, respectively. J.M.M. is supported by the
Jet Propulsion Laboratory, California Institute of Technology, under
contract with NASA. The NRAO is a facility of the National Science
Foundation operated under cooperative agreement by Associated
Universities, Inc.  The Owens Valley Millimeter Array is a radio telescope
facility operated by the California Institute of Technology and is
supported by NSF grants AST 93--14079 and AST 96--13717.  This research
has made use of the NASA/IPAC Extragalactic Database (NED) which is
operated by the Jet Propulsion Laboratory.

\vfill\eject
 
\centerline{Figure Captions}
 
\vskip 0.3in

\noindent
Figure 1.  U$^{\prime}$, B, and I-band images of 3C 293. The U, B, and
I-band images are displayed as blue, green, and red, respectively.  The
images are shown with linear (a) and logarithmic (b) stretches.  North is
up, and east is to the left.

\vskip 0.1in
\noindent
Figure 2.  NRAO 12m CO(1$\to$0) spectrum of 3C{\ts}293. The spectrum is
plotted in units of main beam brightness temperature. The spectrum has
a 1$\sigma$ rms of 0.52 mK per velocity channel, where the resolution of
a single channel is $\sim 43$ km s$^{-1}$.

\vskip 0.1in
\noindent
Figure 3.  OVRO map of the CO(1$\to$0) emission and 3 mm continuum
emission in 3C{\ts}293.  The continuum emission corresponds to a position
of $13^{h}50^{m}03.20^{s}~~+31^{\rm o}41^{\prime}32.8^{\prime \prime}$
(B1950.0).  The 1$\sigma$ rms is 0.00436 and 0.00335 Jy/beam for the high
and low+high resolution data, respectively.  The high resolution data are
plotted as 60, 70, 80, 90, and 99\% contours, where 60\% corresponds to 
3$\sigma$ rms and 100\% corresponds to a peak flux of 0.0218
Jy/beam.  The low+high resolution data are plotted as 50, 60, 70, 80, 90,
and 99\% contours, where 50\% corresponds to 3$\sigma$ rms and
100\% corresponds to a peak flux of 0.0201 Jy/beam.  Spectra extracted at
several positions in the CO emission show clear evidence of rotation, as
well as absorption associated with the central continuum source. The
spectra have a 1$\sigma$ rms of 0.021 Jy per velocity channel, where the
resolution of a single channel is $\sim10$ km s$^{-1}$.  A low-resolution
CO map of 3C 293 can be found in Evans (1998).

\vskip 0.1in
\noindent
Figure 4.  Radio (1.5 GHz) image of 3C{\ts}293 superimposed on the
logarithmic near-infrared (2.17 $\mu$m) image.  The near-infrared image
has been boxcar smoothed 4$\times$4 pixels; the unsmoothed pixel scale is
0.19$\arcsec$/pixel. (a) High resolution 5 GHz MERLIN image and
high-resolution CO(1$\to$0) distribution (see Figure 3) superimposed on
the 7000\AA$~$ HST image.  The pixel scale of the 7000\AA$~$ image is
0.046$\arcsec$/pixel.  North is up, and east is to the left.

\vfill\eject
\begin{deluxetable}{lll}
\pagestyle{empty}
\tablewidth{0pt}
\tablecaption{Properties of CO-Luminous Radio Galaxies}
\tablehead{
\multicolumn{1}{c}{Parameter} &
\multicolumn{1}{c}{3C 293} &
\multicolumn{1}{c}{Other CO-Luminous}\nl
\multicolumn{1}{c}{} &
\multicolumn{1}{c}{} &
\multicolumn{1}{l}{Radio Galaxies}\nl}
\startdata
$D_{\rm L}$ (Mpc) & 180 & 30--500 \nl
$L_{\rm ir}(8-1000\mu$m) ($\times10^{9}$ L$_\odot$) & 33&0.9--1700 \nl
$z_{\rm CO}$ & 0.0446& 0.0018--0.12\nl
$S_{\rm CO}\Delta v$ (Jy km s$^{-1}$) & $51\pm 11$& 30-200\nl
$\Delta v_{\rm FWHM}$ (km s$^{-1}$) & 400 & 250--400\nl
$L^{\prime}_{\rm CO}$ 
($\times10^8$ K km s$^{-1}$ pc$^2$) & 3.7 & 2.5-12 \nl
M(H$_2$) ($\times10^{9}$ M$_\odot$) & 15 & 1--50\nl
$L_{\rm ir} / L^{\prime}_{\rm CO}$ 
($L_\odot/$K km s$^{-1}$ pc$^2$) & 8.9 & 16--220 \nl
\enddata
\tablerefs{Properties of Additional Radio Galaxies: (1) Phillips et al.
1987; (2) Mirabel, Sanders, \& Kaz\`{e}s 1989;
(3) Mazzarella et al. 1993; (4) Evans et al. 1999.}
\end{deluxetable}

\end{document}